\documentstyle[12pt]{article}

    \newtheorem{Definition}{Definition}
        \newenvironment{definition}
            {\begin{Definition}\em}{\end{Definition}}
    \newtheorem{Example}{Example}
        \newenvironment{example}
            {\begin{Example}\em}{\end{Example}}
    \newtheorem{lemma}[Definition]{Lemma}
    \newtheorem{proposition}[Definition]{Proposition}

    \newtheorem{Remark}{Remark}

    \newtheorem{Acknowledgements}{Acknowledgements}

    \newtheorem{Acknowledgement}{Acknowledgement}
        
        \newenvironment{acknowledgement}{\begin{Acknowledgement}\em}
            {\end{Acknowledgement}}

\newcommand{\qed}{\text{\rule{.4em}{1.7ex}\hspace{.6em}}}
\newenvironment{proof}{\noindent {\bf Proof:\ }}{\hspace*{.1em}\hfill\qed
\bigskip \noindent}
\newcommand{\suchthat}{|\hspace{.2em}}
\newcounter{rom}
{\end{list}}

\newcounter{abc}
\newenvironment{abclist}{\begin{list}{(\alph{abc})}
        {\setlength{\leftmargin}{2em}\usecounter{abc}}}%
{\end{list}}
\newcommand{\R}{{\Bbb R}}
\newcommand{\N}{{\Bbb N}}
\newcommand{\Ric}{{\operatorname{Ric}}}

\newcommand{\grad}{{\operatorname{grad}}}
\newcommand{\nab}[2]{\nabla\raisebox{-.8ex}{$#1$}#2}

\renewcommand{\d}{\partial}

\newcommand{\eqref}[1]{{(\ref{#1})}}

\newcommand{\sign}{{\operatorname{sign}}}
\newcommand{\SPAN}{{\operatorname{span}}}
\newcommand{\lab}[1]{\label{#1}}
\newcommand{\Rad}{{\operatorname{Rad}}}
\newcommand{\x}{{\hat{x}}}
\newcommand{\D}{{\cal D}}

\newcommand{\tx}{{\tilde{x}}}
\newcommand{\ttt}{{\tilde{t}}}
\newcommand{\tg}{{\tilde{g}}}

\newcommand{\tD}{{\tilde{\cal D}}}

\newcommand{\htt}{{\hat{t}}}
\newcommand{\hg}{{\hat{g}}}
\newcommand{\hphi}{{\hat{\phi}}}
\newcommand{\hT}{{\hat{T}}}
\newcommand{\hx}{{\hat{x}}}
\newcommand{\ML}{{\tilde M}^{-}}
\newcommand{\MR}{{\tilde M}^{+}}
\newcommand{\txarray}{{\tx^1,\tx^2,\tx^3}}
\newcommand{\xarray}{{x^1,x^2,x^3}}
\newcommand{\hxarray}{{\hx^1,\hx^2,\hx^3}}

\newcommand{\tM}{{\tilde M}}
\newcommand{\GG}{{\cal G}}

\newenvironment{align*}{\[\begin{array}{ll}}{\end{array}\]}
                    {\end{array}\end{equation}}

\newenvironment{equation*}{\[}{\]}
\newenvironment{split}{\begin{array}{ll}}{\end{array}}
                    {\end{eqnarray}}
\newenvironment{multline*}{\[\begin{array}{l}}{\end{array}\]}

\newcommand{\binom}[2]{(#1,#2)}
\newcommand{\text}[1]{\mbox{#1}}
\newcommand{\operatorname}[1]{\mbox{#1}}
\renewcommand{\R}{\mbox{$\rm I\! R$}}
\renewcommand{\N}{\mbox{$\rm I\! N$}}
\renewcommand{\Ric}{{\mbox{Ric}}}

\renewcommand{\grad}{{\mbox{grad}}}
\renewcommand{\nab}[2]{\nabla\raisebox{-.8ex}{$#1$}#2}

\renewcommand{\sign}{{\mbox{sign}}}

\begin{document}
\setcounter{page}{1}
\title{Black holes, cosmological singularities and change of signature}
\author{Marcus Kriele \\
\mbox{\small Fachbereich Mathematik, Sekr. MA 8-3} \\
\mbox{\small Technische Universit\"{a}t Berlin} \\
\mbox{\small Strasse des 17 Juni 136, 10623 Berlin} \\
\mbox{\small Federal Republic of Germany} \\
\\ J\'er\^ome Martin$^*$ \\
\mbox{\small Department of Applied Mathematics and Theoretical Physics} \\
\mbox{\small University of Cambridge, Silver street} \\
\mbox{\small Cambridge CB3 9EW, United Kingdom}}
\maketitle
\begin{abstract}
There exists a widespread belief that signature type change could be
used to avoid spacetime singularities.  We show that signature change
cannot be utilised to this end unless the Einstein equation is
abandoned at the suface of signature type change.  We also discuss how
to solve the initial value problem and show to which extent smooth and
discontinuous signature changing solutions are equivalent.
\end{abstract}
\footnotesize
${}^*$ Permanent adress: Laboratoire de Gravitation
et Cosmologie Relativistes, Universit\'e Pierre et Marie Curie,
CNRS/URA 769, Tour 22/12, Boite courier 142, 4 place Jussieu 75252 Paris
cedex 05, France.
\normalsize

\eject

\setcounter{page}{2}
\section{Introduction}
According to the Hawking-Penrose theorem \cite{hawking-penrose-70a},
singularities in General Relativity seem to be unavoidable. The two
most well-known examples are the singularities of the
Friedmann-Robertson-Walker metric ("big bang singularities") and of
the Schwarzschild solution ("black hole singularities").  The presence
of these singularities is usually interpreted as the sign that General
Relativity, a classical theory, is no more valid since quantum effects
have to be taken into account when the curvature reaches the Planck
limit. Therefore, it is not surprising that a possible solution to
this problem has been suggested in the context of quantum cosmology.
Recent studies
\cite{hayward-92a,ellis-sumeruk+92a,dereli-tucker-93a,hayward-93-p-b,kossowski-kriele-93a,kossowski-kriele-94b,kossowski-kriele-94a,kerner-martin-93a,martin-94a,hellaby-dray-94a,kossowski-kriele-93b}
have shown that change of signature can be also a feature of classical
General Relativity.  In this framework, the very early Universe is
described by a Riemannian\footnote{In the physics literature the term
`Euclidean' is often used instead of the term `Riemannian' which is
more common in the mathematical literature.} manifold which does not
have a big bang singularity \cite{hartle-hawking-83a}. It has been hoped that
this is a
consequence of signature change (this has been argued in
\cite{ellis-sumeruk+92a} since the singularity theorems do not
apply for Riemannian manifolds).  However, in this paper we will show
that big bang singularities which would occur without signature change
will reappear as Big Crunch singularities.  We also answer the
question whether one can employ signature change in order to avoid
black hole singularities to the negative.

There have been put forward different suggestions as to how
to implement signature change classically.  As a consequence, there
are now different competing theories and an ongoing discussion about
the relative merit of the smooth and discontinuous description of
signature type change
(cf.
\cite{hayward-92a,ellis-sumeruk+92a,dereli-tucker-93a,hayward-93-p-b,kossowski-kriele-93b,carfora-ellis-94-p-a}).  These
different proposals can be divided into two groups.
\begin{abclist}
\item
One imposes regularity conditions at the hypersurface of signature
change which can be understood as imposing the Einstein equations (in a
suitable form) at the surface of signature type change. This approach
has been adopted by
\cite{hayward-92a,dereli-tucker-93a,hayward-93-p-b,kossowski-kriele-93a,kossowski-kriele-94b,kossowski-kriele-94a,kerner-martin-93a,martin-94a,kossowski-kriele-93b}.
\item
One views spacetime as a 1-parameter family of Riemannian manifolds
and therefore relaxes the regularity conditions at the surface of
signature change.  This approach has been adopted by
\cite{ellis-sumeruk+92a,carfora-ellis-94-p-a}.
\end{abclist}
We are of the opinion that wherever one can use the Einstein equation
one should impose it and are therefore favouring approach (a) which we
will consider exclusively. Within approach (a) there are two competing
proposals: One can implement signature change with a discontinuous but
non-degenerate or with a continuous but degenerate metric.  So
naturally the question arises whether any of these two implementations
is superior.  In \cite{kossowski-kriele-93b} this question has been
atttempted to decide from a geometrical viewpoint.  The authors
concluded that the smooth description was vastly superior.  However,
this conclusion rests an ``a priori'' demands on the regularity of the
solutions.  Here we show that for solutions of Einstein's equations
different regularity conditions arise naturally in the discontinuous
description.  Assuming these regularity conditions the space of
solutions of the Einstein equations in either scenario are canonically
equivalent.  Thus it appears to be a matter of taste which setting one
prefers.

Our paper is organized as follows: In section \ref{s2} we discuss the
initial value problem and show the equivalence between continuous and
discontinuous change of signature for a specific class of solutions.
In section \ref{s3.1} we show that big bang singularities of Lorentzian
solutions
will reappear as Big Crunch singularities in the corresponding type
changing solution.  In section \ref{s3.2} we prove the impossibility of
matching an Riemannian manifold inside the black hole horizon without
introducing new singularities.

\section{Comparison between continuous and discontinuous change of
signature} \lab{s2}

Let us first recall the two definitions.  In the discontinuous case
one should restrict to signature change at {\em spacelike\/}
hypersurfaces $\tD$ because only in this case it is possible for $\tD$
to inherit the same structure from both the Lorentzian and the
Riemannian region.  But given a distinguished spacelike hypersurface
one can define a distinguished time function, the parameter
function of the unit geodesics starting orthogonal to this
hypersurface. To employ this natural time function has several
advantages.  For instance, it makes it possible to write down the
energy momentum tensor as a well defined object and it  facilitates the
comparison of smooth and discontinuous signature change.  We do not
want to specify the regulariy of the considered type changing metric
yet.  So let  $\GG$ be a subset of all
functions $f: M \rightarrow \R$ to be specified later and define:

\begin{definition}                                      \lab{d1}
$(\tM,\tg)$ is a {\em type changing spacetime with jump of class $\GG$
 \/} if $\tM$ is a smooth, 4-dimensional manifold with an everywhere
 non-degenerate, symmetric $\binom{0}{2}$ tensor field $\tg$ which is
 continuous everywhere except at a hypersurface $\tD$. For any $x$ in
 $\tD$ there exists a coordinate neighbourhood such that $\tg$ is given
 by $\tg = - \eta d\ttt^2+\tg_{ij}(\ttt,\txarray)d\tx^i\d\tx^j ,$
where $\eta = \sign(\ttt)$ and $\tg_{ij} \in \GG$  $(i,j \in \{1,2,3\})$.
\end{definition}

In \cite{kossowski-kriele-93b} the class $\GG $ has been taken as the
class of $C^k$-functions.  This has been justified since the
$\ttt$-coordinate is the invariantly defined time function so that one
can view the $\tg_{ij}$ as a $C^k$-differentiable 1-parameter family
of $C^k$-3-metrics.  However, this choice of class is not natural from
a physical point of view as we will see below.

For the smooth case one may define \cite{kossowski-kriele-93a}:

\begin{definition}                                      \lab{d2}
$(M,g)$ is a {\em transverse, type changing $C^k$-spacetime\/} ($k \in
\N\cup\{0,\infty, \omega\}$)\footnote{A function $f$ is said to be $C^
{\omega }$ if $f$ is real analytic.} if $M$ is a smooth, 4-dimensional
manifold with a symmetric $C^k$-$\binom{0}{2}$-tensor field $g$ such
that at any point $x\in \D := \{ \x\in M\suchthat g_{|\x} \text{ is
degenerate}\}$ we have
$d\left(\det(g_{ab})\right)_{|x} \neq 0$ for some (and hence any)
coordinate system.
Moreover, at any point where $g$ is non-degenerate it is either
Riemannian or Lorentzian.
\end{definition}

The main difference to definition \ref{d1} is that here $g$ is assumed
to be a smooth tensor field.  It is clear that the surface of
signature change must be given by $\det(g_{ab}) = 0$.  The additional
condition, $d\left(\det(g_{ab})\right)_{|x} \neq 0$, may be thought of
as a  genericity condition.

This definition also allows to have signature change at null surfaces.
  However, for cosmological applications one would like to have a
  spacelike surface of signature change.  Thus we define

\begin{definition}      \lab{d3}
$(M,g)$ has a {\em transverse radical\/} if in addition $\Rad_{x} :=
\{ v_x \in T_xM\suchthat g(v_x,\cdot) = 0\}$ intersects $T_x\D$
transversely for all $x\in\D$.
\end{definition}

Observe that $\Rad_x$ is necessarily one-dimensional.
Notice that since $g$ is a well defined tensor field it is natural to
consider $C^k$ metrics $g$.  In contrast to the discontinuous case we
do not need to specify an (invariantly defined) system of
coordinates.  We will see below that the class of such metrics is
also  natural from a physical point of view.

Although these definitions are rather different in spirit one can
introduce coordinates $(t,\xarray)$ such that the content of
definitions \ref{d2}, \ref{d3} can be reformulated similarly to definition
\ref{d1}:

\begin{lemma}   \lab{t3}
Let $M$ be a manifold, $g$ be a $\binom{0}{2}$-tensor field, and $\D
:= \{ \x \in M \suchthat g_{\x}$ is degenerate$\}$.  Then
$(M,g)$ is a transverse, type changing spacetime with transverse
radical if and only if for any $\x
\in \D$ there exists a neighbourhood and coordinates $(t,\xarray)$
such that
\[
g = -tdt^2+g_{ij}(t,\xarray)dx^idx^j.
\]
\end{lemma}

\begin{proof}
This has been shown (in a more general context) in
\cite{kossowski-kriele-94a}.
\end{proof}

It is now possible to relate the smooth picture to the discontinuous
one.  A necessary condition for equivalence is clearly that the
Lorentzian and the Riemannian parts of the two descriptions are
isometric. Thus two metrics $g,\tg$ are {\em equivalent\/} if there
exists a homeomorphism which is an isometry away from the surfaces of
signature type change. Since the surfaces $\ttt$ = const and $t =$
const have an invariant meaning such a transformation must be given by
\[
\psi\colon (t,\xarray)  \mapsto  (\ttt,\txarray) \text{ with } \ttt =
\eta \frac{2}{3}\left(\eta t \right)^{3/2}, \quad \tx^i = \tx^i(\txarray).
\]
It is clear that we only need to consider  transformations
$\psi$ with $\tx^i = x^i$.

Then $(M,\tg)$ is equivalent to a metric transverse, type changing
$C^k$-spacetime if and only if $(M,\psi^*\tg)$ satisfies definition
\ref{d2}.

We will now consider the Einstein equations and show that the solutions
in both the discontinuous and the smooth approach are canonically
equivalent. Assume for definiteness\footnote{But compare the
conclusion in which we point out a key feature of the scalar field
energy momentum tensor which makes the following discussion possible}
that the energy momentum tensor has the form of a scalar field with
arbitrary potential (but which is not coupled to the scalar
curvature):
\[
T_{ab} = \d\phi_a\d\phi_b -\frac{1}{2}\left(
g(\grad(\phi),\grad(\phi))+V(\phi)\right)g_{ab},
\]
where $V$ is some analytic function and $\phi$ is the scalar field.
For convenience we will also assume that $M\setminus\D$ consists of two
connected components, a Riemannian component $\MR$ and a Lorentzian
component $\ML$.

In the discontinuous case (and arbitrary coordinates) the Einstein
equations will in general fail to be distributional.  Even in our
adapted (shift = 0) coordinates $(\ttt, \txarray)$ the component of
the Ricci tensor will in general fail to be defined
distributionally. However, the energy momentum tensor is a well
defined distributional tensor and it makes sense to demand that it is
bounded.  This requirement just means that at the surface of signature
type change we do not have a singularity due to concentration of
matter.  From this condition it follows that $\d_\ttt g_{ij} = 0$ at
$\ttt = 0$ and that {\em then\/} also the components of the Ricci
tensor are defined distributionally\footnote{The above follows
immediately from the expressions in the appendix of
\cite{kossowski-kriele-93b}.  Observe that a similar but less
convoluted claim on the bottom of page 2366 in this paper is not true
(this, however, does not affect the rest of the paper)}.

The initial value problem at the surface of signature change splits
into qualitatively different initial value problems, one for the
Lorentzian and the other one for the Riemannian part.

Let us first consider the Lorentzian part.  We will denote all
quantities with a hat in order to distinguish from the signature
changing case. The condition that
$\hT_{ab}$ is bounded was equivalent to the requirement $\d_\htt
\hg_{ij} = 0$ for the inital 3-metric.  We therefore have only to
solve the usual Einstein equation for this sort of initial data.  This
is (given smooth data) always possible if the usual constraint
equation holds at the surface.  The proof of Lemma 2 in
\cite{kossowski-kriele-94b} implies in addition that the Taylor series
of $\hg_{ij}$ and of the scalar field $\hphi$ depend smoothly on
$\htt^2$.  Observe that in the analytic\footnote{I.e., each metric
components $\hg_{ij}$ can be expressed as a power series in the
variables $(\htt,\hxarray)$} case at $\htt = 0$ we not only have a
surface of infinetesimal time symmetry but that the map $\htt
\rightarrow -\htt$ induces an isometry.

For the Riemannian part it would be much more difficult to solve the
initial value problem because the system of differential equations is
not hyperbolic in this region.  But one can construct a Riemannian
solution from a Lorentzian solution employing the trick with the Wick
rotation: If one replaces $ \htt $ with $\check t = i\htt$ in the
Lorentzian solution one automatically obtains a solution of the
Riemannian equations.  Moreover, this Riemannian solution has a real
Taylor series and therefore is real if it is analytic (cf.  appendix
\ref{sa} for an illustration that one {\em must\/} impose
analyticity).

It is now clear how to obtain a signature changing solution $(M,\tg)$:
We use the Wick rotation in one connected component of $M\setminus\D$
which we call $\MR$.  Observe that the metric coefficients now depend
analytically on $\eta\ttt^2$.  Thus $\GG$ should be assumed to be the
class of functions which depend analytically on $\eta
\ttt^2,\txarray$.

In order to obtain a smooth signature type changing solution we now
apply the transformation $\psi$ to the discontinuous solution.  This
is possible since the discontinuous solution depends analytically on
$\eta\ttt^2$. This transformation results in an analytic solution
which depends analytically on $t^3$.  Moreover, any analytic,
transverse, type changing solution with bounded energy momentum tensor
can be obtained in this way \cite{kossowski-kriele-94b}.

We have therefore established the following diagram:

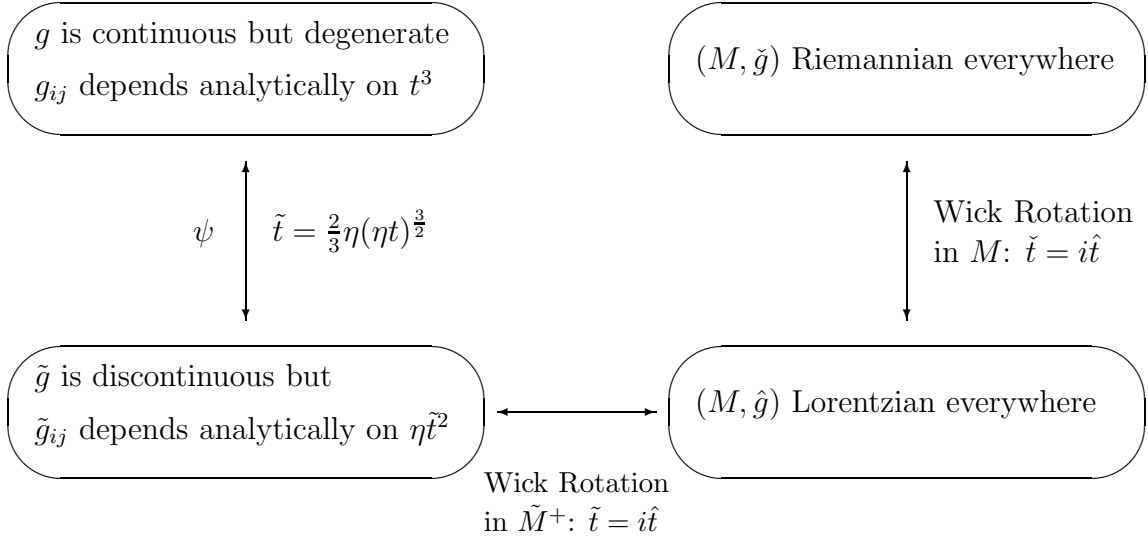
\begin{picture}(500,300)(-25,0)

\put(50,100){\oval(180,50)}
\put(50,230){\oval(180,50)}
\put(300,100){\oval(180,50)}
\put(300,230){\oval(180,50)}

\put(175,100){\vector(-1,0){30}}
\put(175,100){\vector(1,0){30}}
\put(50,165){\vector(0,-1){30}}
\put(50,165){\vector(0,1){30}}
\put(300,165){\vector(0,-1){30}}
\put(300,165){\vector(0,1){30}}

\put(30,165){$\psi $}
\put(60,165){$\tilde{t}=\frac{2}{3}\eta (\eta t)^{\frac{3}{2}}$}
\put(140,70){{\small Wick Rotation}}
\put(140,55){{\small in $\MR$: $\tilde{t}=i\hat{t}$}}
\put(310,172){Wick Rotation}
\put(310,157){in $M$: $\check t = i \hat t$}

\put(-30,240){$g$ is continuous but degenerate}
\put(-30,220){$g_{ij}$ depends analytically on $t^3$}
\put(-30,110){$\tilde{g}$ is discontinuous but}
\put(-30,90){$\tilde{g}_{ij}$ depends analytically on $\eta\tilde{t}^2$}
\put(220,100){$(M,\hat{g})$ Lorentzian everywhere}
\put(220,90){}
\put(220,230){$(M,\check{g})$ Riemannian  everywhere}
\put(220,220){}

\end{picture}

\begin{example}		\lab{e1}
In order to illustrate the previous diagram, let us consider the
Friedmann-Robertson-Walker (FRW) metric with a cosmological
constant (this example has been already studied in
\cite{hayward-92a,ellis-sumeruk+92a}).  In the case where the
continuous choice is made, the metric can be written as:

\[
ds^2 = \left\{
\begin{split}
-t{d}t^2+\cos ^2(\frac{2}{3}(-t)^{\frac{3}{2}}){d}\Omega ^2_3
&\qquad -(\frac{3\pi }{4})^{\frac{2}{3}} \leq t \leq 0 \\
-t{d}t^2+\cosh ^2(\frac{2}{3}t^{\frac{3}{2}}){d}\Omega ^2_3
& \qquad t\geq 0
\end{split}
\right.
\]

\noindent
where ${\rm d}\Omega ^2_3$ represents the line element of a
three-sphere. The metric is manifestly degenerate at the surface $t=0$ but
is $C^{\infty }$ since the Taylor series of $\cos (\frac{2}{3}(-t)^
{\frac{3}{2}})$ and $\cosh (\frac{2}{3}t^{\frac{3}{2}})$ are the same:
\begin{eqnarray}
\cos (\frac{2}{3}(-t)^{\frac{3}{2}})=\cosh (\frac{2}{3}t^{\frac{3}{2}})=
1+\frac{1}{2!} \left(\frac{2}{3}\right)^2t^3+\frac{1}{4!}
\left(\frac{2}{3}\right)^4 t^6+\cdots
\nonumber
\end{eqnarray}
{}From this expression, it is obvious that $g_{ij}$ depends analytically
on $t^3$.
\par
We can now perform the transformation $\tilde{t}=\frac{2}{3}
\eta (\eta t)^{\frac{3}{2}}$, $\tilde{x}_i=x_i$; then the metric takes
the form:

\[
d{\tilde s}^2 = \left\{
\begin{split}
d\tilde{t}^2+\cos ^2(\tilde{t}){d}\Omega ^2_3 &\qquad -(\frac{\pi }{2}) \leq
\tilde{t} \leq 0 \\
-{d}\tilde{t}^2+\cosh ^2(\tilde{t}){\rm d}\Omega^2_3
&\qquad\tilde{t} \geq 0
\end{split}
\right.
\]

The metric is no more degenerate on the surface $\tilde{t}=0$ but is
discontinuous. $\tilde{M}^+$ is half of the sphere $S^4$ whereas
$\tilde{M}^-$ is half of the De Sitter spacetime. $\tilde{g}_{ij}$ depends
analytically on $\eta \tilde{t}^2$.
\par
If we use the Wick rotation $\tilde{t}=i\hat{t}$ on $\tilde{M}^+$, we obtain
the following metric on $M$:
\begin{eqnarray}
{\rm d}s^2 = -{\rm d}\hat{t}^2+\cosh ^2(\hat{t}){\rm d}\Omega ^2_3.
\nonumber
\end{eqnarray}
$M$ is now the entire De Sitter spacetime.
\par
Finally, if we perform a Wick rotation on $(M,\hat{g})$, the metric becomes:
\begin{eqnarray}
{\rm d}s^2 = {\rm d}\check{t}^2+\cos ^2(\check{t}){\rm d}\Omega ^2_3,
\nonumber
\end{eqnarray}
namely the metric of the sphere $S^4$. Then, all the relationships of the
diagram have been explicitely shown using the simple example of the FRW
metric.
\end{example}

\section{Spacetime singularities} \lab{s3}
\subsection{Signature change in cosmology} \lab{s3.1}

We show that one cannot use signature type change in order to turn singular,
inextensible spacetimes into non-singular ones:

\begin{proposition}
Let $(M,g)$ be a signature type changing spacetime.  If it is
singularity-free then the Lorentzian spacetime $(M,\hg)$ corresponding
to it is singularity free.
\end{proposition}

\begin{proof}
The discussion in section \ref{s2} shows that the Lorentzian spacetime
$(M,\hg)$ corresponding to $(M,g)$ admits a reflectional isometry
$\htt \mapsto -\htt$.  Hence if it has a singularity in $\MR$ then it
also must have a singularity in $\ML$.  Since $\ML$ is isometric to one
connected component of $M\setminus \D$ in the Lorenztian solution our
claim follows.
\end{proof}

Actually, there is also a different argument first pointed out by
Hayward \cite{hayward-93a}.  Since the surface of signature change is
totally geodesic in a closed cosmology the universe is immediately
contracting unless one violates the strong energy condition.  Hayward
interpreted this circumstance as positive evidence for an inflationary
phase in which the strong energy condition would be violated.

Notice that in example \ref{e1} above all solutions are
singularity-free.

\subsection{Signature change at black holes}\lab{s3.2}

One may speculate that it could be possible to avoid black hole
singularities by imposing change of signature at the boundary of the
black hole.  In this section we show that a signature changing
mechanism would not work.  Recall that there are two possible
definitions as to what the boundary of a black hole is:
\begin{abclist}
\item
{}From a quasi-local point of few it is natural to consider the outer
trapping  horizon as the boundary of the black hole
\cite{hayward-94-p-a}.
\item
One may also take the event horizon of a black hole as a definition of
its boundary. However, this definition is essentially global and
therefore does not capture the physical content as well as the
definition in (a).
\end{abclist}

\noindent
An outer trapping horizon is a hypersurface surface which is foliated
by outer marginally trapped surfaces (outer means that the expansion
of the null direction which vanishes would become negative when the
surface is moved into the other null direction).  Hayward has also
shown that the outer trapping horizon is spacelike if the null energy
condition and a genericity condition hold \cite[Theorem
2]{hayward-94-p-a}. Thus signature change at the trapping horizon
would be mathematically equivalent to cosmological signature change.
In particular, if the energy momentum tensor is supposed to be bounded
then one has to assume that this surface is a surface of
(infinetesimal) time symmetry.  This is certainly impossible where a
physical black hole develops.  Thus one cannot implement signature
type change at the trapping horizon.

It follows that signature change could only be implemented at the
event horizon of a black hole.  Observe that the event horizon is a
null surface $\D$ and that the weakest regularity condition to impose
on $\D$ is that the induced metric is unambigously defined (ie. the
metric inherited from the Lorentzian part should be the same as the
metric inherited from the Riemannian part). Since in Riemannian
geometry there do not exist non-vanishing null vectors it is clear
that discontinuous signature type change is impossible.

However, continuous signature type change is not ruled out yet.  As a
trivial example consider the 2-dimensional metric $ydx^2+dy^2$.  This
is a transverse, type changing metric for which the surface of
signature change $\D$ is given by $y = 0$.  However, the radical,
$\SPAN\{\d_x\}$ is tangent to $\D$ and hence it fails to be
transverse. Thus $\D$ is a null surface.  But if one calculates the
Gau\ss\ curvature $K$  one obtains $K = 1/y^2$ which diverges at $y = 0$

We will now show that signature change at a null hypersurface implies
the existence of curvature singularities.  If $\D$ is null then the
radical $\Rad$ must be everywhere tangent to $\D$. Thus it suffices to
prove the following proposition which is an extension of
\cite[Theorem 3]{kossowski-kriele-93-p-a}:

\begin{proposition} \lab{p4}
If $(M,g)$ is a transverse type changing spacetime but the radical not
transverse at $x \in \D$ then the energy momentum tensor $T_{ab}$ is
unbounded at $x$.
\end{proposition}

\begin{proof}
There exists an adapted orthonormal frame $\{e_0,e_1,e_2,e_3\}$
such that $g(e_i,e_j) = \delta^i_j$, $g(e_0,e_i) = 0$ for $i,j \in
\{1,2,3\}$ and $g(e_0,e_0) = \tau$ where $\tau$ is a function with
$\tau_{|\D} = 0$ but $d\tau_x \neq 0$.  The existence of such a frame
follows with a slight adaptation of the Gram-Schmidt procedure
\cite{kossowski-85a}.   Since $g(\nab{e_a}e_b,e_c)$ extends
smoothly and since
\begin{align*}
g(\nab{e_a}\nab{e_b}e_a,e_b) & =
\nab{e_a}g(\nab{e_b}e_a,e_b)-g(\nab{e_b}e_a,\nab{e_a}e_b) \\
 & = -\frac{1}{\tau}g(\nab{e_b}e_a,e_0) g(e_0,\nab{e_a}e_b) + \text{
smooth terms},
\end{align*}
we get
\begin{align*}
g(R(e_a,e_b)e_a,e_b) &= -\frac{1}{\tau}
\left(g(\nab{e_b}e_a,e_0) g(\nab{e_a}e_b,e_0) -
 g(\nab{e_a}e_a,e_0) g(\nab{e_b}e_b,e_0)\right)\\
& \qquad + \text{smooth terms}.
\end{align*}

Setting $e_b = e_0$, observing that at $x$ we have $0 = d\tau(e_0) = 2
g(\nab{e_0}e_0,e_0)$, and using that $g([e_a,e_0],e_0)$ vanishes at
$\D$, we obtain
\[
g(R(e_a,e_0)e_a,e_0) = -\frac{1}{\tau}
\left( g(\nab{e_a}e_0,e_0)\right)^2+ \text{smooth terms}.
\]
Now notice that the frame $\{e_0,e_1,e_2,e_3\}$ can be chosen so that
$e_2, e_3$ are tangent to $\D$.  Thus at $x$ we have $d\tau(e_0) =
d\tau(e_2) = d\tau(e_3) = 0$ but $d\tau(e_1) \neq 0$.  Since $
g(\nab{e_a}e_0,e_0) = \frac{1}{2} d\tau(e_a)$ we conclude that
$g(R(e_1,e_0)e_1,e_0)$ diverges (such that $\tau g(R(e_1,e_0)e_1,e_0)$
is bounded) whereas $g(R(e_a,e_0)e_a,e_0)$ is bounded for $a = 0,2,3$.
In particular, $\Ric(e_0,e_0)$ diverges such that $\tau \Ric(e_0,e_0)$
and $(\Ric(e_0,e_0)-g(R(e_1,e_0)e_1,e_0))/\tau$ are bounded.  For
$\alpha \in \{2,3\}$ we have
\[
\Ric(e_{\alpha},e_{\alpha}) =
\sum_{b=0}^3 (g(e_b,e_b))^{-1}g(R(e_{\alpha},e_b)e_{\alpha},e_b).
\]
$\tau\Ric(e_{\alpha},e_{\alpha})$ is bounded since
$g(R(e_{\alpha},e_0)e_{\alpha},e_0)$,
$g(R(e_{\alpha},e_b)e_{\alpha},e_b)$, $\tau
g(R(e_{\alpha},e_b)e_{\alpha},e_b)$, $(g(e_b,e_b))^{-1}$ are bounded
for $b \neq 0$.
\begin{eqnarray*}
\Ric(e_1,e_1) & = &
\sum_{b=0}^3 (g(e_b,e_b))^{-1}g(R(e_1,e_b)e_1,e_b) \\
& = & \sum_{\alpha=2}^3
(g(e_\alpha,e_\alpha))^{-1}g(R(e_1,e_\alpha)e_1,e_\alpha) +
\frac{1}{\tau}
\left(\Ric(e_0,e_0)+\text{smooth terms}\right).
\end{eqnarray*}
Thus we obtain for the scalar curvature $s =
\frac{2}{\tau}\left(\Ric(e_0,e_0) + \text{ smooth terms}\right)$.  In
particular it follows that
$T(e_2,e_2) = -\frac{1}{\tau}\Ric(e_0,e_0) +\frac{1}{\tau}\left(\text{
smooth terms}\right)$ diverges at a similar rate as $\frac{1}{\tau^2}$.
\end{proof}

Observe that the singularity provided by proposition \ref{p4} is
non-distributional since $\tau^2 T(e_2,e_2)$ does not converge to
zero.  This  would be different if $\tau T(e_2,e_2)$ was bounded.

\section{Conclusion}

\noindent
In section \ref{s2} we have shown that there exists a well defined
equivalence of solutions of Einsteins equations for both the
continuous and the discontinuous implementation of signature change.
In order to establish this equivalence we have used an energy momentum
tensor for a scalar field.  The form of this energy momentum tensor is
crucial for the argument.  The key feature is that time derivatives
occur only quadratically ($\d_\ttt a \d_\ttt b)$ and that each such
quadratic pair is weighted with the $\ttt\ttt$-component of the
inverse of the metric.  This circumstance is responsible for the fact
that the Taylor series of the $\tg_{ij}$ depend quadratically on
$\ttt$.  If this was not the case the Wick transformation could not
give rise to real solutions.  We doubt wether there exist any
solutions for energy momentum tensors which do not have this property.
In principle, it is possible to construct Lagrangians whose energy
momentum tensor violates this property.  For instance, one may have as
matter quantities a scalar field $\nu$ and a vector field $X$ and add
a term $d\nu(X)$ to the Lagrangian.

We have established that it is a matter of taste whether one prefers
continuous or discontinuous signature change, provided Einstein's
equations are imposed.  If one works in the discontuinuous picture one
has the advantage that the metric is nowhere degenerate and that there
exist observer fields which can be smoothly continued into the
Riemannian region.  However, one has to pay the price that solutions
of Einstein's equation have unusual regularity properties.   For
instance, due to the factor $\eta$ in
general the energy momentum tensor fails to be analytic
 unless spacetime is static \cite{kossowski-kriele-93b}.   On the other
hand, observe that the  transformation $\psi$ transforms any
such solution into an analytic transverse type changing solution for
the analogous problem with an analytic energy momentum tensor.  Thus,
from a mathematical point of view, the continuous picture seems to be
more familiar.

Finally, we have shown that signature change cannot be sucessfully
employed in order to avoid singularities of solutions of Einstein's
equations.

\appendix
\section{Appendix}\lab{sa}
Consider a smooth but non-analytic Lorentzian solution wich depends on
$\ttt^2$.  Applying the Wick rotation we obtain a Riemannian solution
of the corresponding Riemannian Einstein equations.  This Riemannian
solution has a real Taylor series.  Nevertheless, this solution fails
to be real in general. In order to illustrate this point consider the
simple but completely analogous massless wave equation on a
flat, 2-dimensional background.  In the Lorentzian region it is the
usual wave equation $-\d_\ttt\d_\ttt\phi +
\d_\tx\d_\tx\phi = 0$ whereas in the Riemannian part it is the usual
Laplace equation $\d_\ttt\d_\ttt\phi +\d_\tx\d_\tx\phi = 0$, both with
initial condition $\d_\ttt\phi(0,\tx) = 0, \phi(0,\tx) =
\vartheta(\tx)$ for some arbitrary function $\vartheta$.  Although for
any smooth initial function $\vartheta$ the Lorentzian region has a
unique, smooth solution $\phi(\ttt,\tx) = \vartheta(\tx+\ttt) +
\vartheta(\tx-\ttt)$ it is well known that any smooth solution
in the Riemannian region must have {\em analytic\/} initial data
\cite[p. 455]{garabedian-86}.  Of course, we could construct a
solution of the Riemannian equation using Wick rotation.  However,
this solution fails to be real even though  its Taylor series is
automatically real.  As an explicit example, consider  the function
\[
\vartheta(x) :=
\left\{
\begin{split}
 0 &\text { for } x \geq 0\\
x e^{-1/x^2}& \text{ for } x < 0.\end{split}
\right.
\]
The Taylor series of this function centred at $\ttt = 0$ vanishes and
therefore is real.   But
is is clear that $\phi(\ttt,\tx)$ is not a real valued function.

Thus one has to demand that the initial data are analytic.

\begin{acknowledgement}
M. K. would like to thank the Laboratoire de gravitation et cosmologie
relativiste of the Universit\'e Pierre et Marie Curie for warm
hospitality. J. M. would like to thank the Minist\`ere de la Recherche
et de l'Enseignement Sup\'erieur for a research grant.  We would like
to thank Robin Tucker for pointing out that to refer to $\psi$ as a
`coordinate transformation' is misleading.
\end{acknowledgement}

\bigskip

\eject
\bibliographystyle{plain}

\begin{thebibliography}{99}

\bibitem{hawking-penrose-70a}
S.~W. Hawking and R.~Penrose.
\newblock The singularities of gravitational collapse and cosmology.
\newblock {\em Proc. Roy. Soc. Lond. A}, 314:529--548, 1970.

\bibitem{hayward-92a}
S.~A. Hayward.
\newblock Signature change in general relativity.
\newblock {\em Class. Quantum Grav.}, 9:1851--1862, 1992.

\bibitem{ellis-sumeruk+92a}
G.~F.~R. Ellis, A.~Sumeruk, D.~Coule, and C.~Hellaby.
\newblock Change of signature in classical relativity.
\newblock {\em Class. Quantum Grav.}, 9:1535--1554, 1992.

\bibitem{dereli-tucker-93a}
T.~Dereli and R.~W. Tucker.
\newblock Signature dynamics in general relativity.
\newblock {\em Class. Quantum Grav.}, 10:365--373, 1993.

\bibitem{hayward-93-p-b}
S.~A. Hayward.
\newblock Junction conditions for signature change.
\newblock {\em Preprint gr-qc/9303034}, 1993.

\bibitem{kossowski-kriele-93a}
M.~Kossowski and M.~Kriele.
\newblock Signature type change and absolute time in general relativity.
\newblock {\em Class. Quantum Grav.}, 10:1157--1164, 1993.

\bibitem{kossowski-kriele-94b}
M.~Kossowski and M.~Kriele.
\newblock The {E}instein equation for signature type changing spacetimes.
\newblock {\em Proc. Roy. Soc. Lond. A},  446:115--126, 1994.

\bibitem{kossowski-kriele-94a}
M.~Kossowski and M.~Kriele.
\newblock Transverse, type changing, pseudo {R}iemannian metrics and the
  extendability of geodesics.
\newblock {\em Proc. Roy. Soc. Lond. A}, 444:297--306, 1994.

\bibitem{kerner-martin-93a}
R.~Kerner and J.~Martin.
\newblock Change of signature and topology in a five-dimensional cosmological
model.
\newblock {\em Class. Quantum Grav.}, 10:2111--2122, 1993.

\bibitem{martin-94a}
J.~Martin.
\newblock Hamiltonian quantization of General Relativity with the change
of signature.
\newblock {\em Phys. Review D}, 49:5086--5095, 1994.

\bibitem{hellaby-dray-94a}
C.~Hellaby and T.~Dray.
\newblock Failure of standard conservation laws at a classical change
of signature.
\newblock {\em Phys. Review D}, 49:5096--5104, 1994.

\bibitem{kossowski-kriele-93b}
M.~Kossowski and M.~Kriele.
\newblock Smooth and discontinuous signature type change in general relativity.
\newblock {\em Class. Quantum Grav.}, 10:2363--2371, 1993.

\bibitem{hartle-hawking-83a}
J.~B. Hartle and S.~W. Hawking.
\newblock Wave function of the universe.
\newblock {\em Phys. Review D}, 28(12):2960--2975, 1983.

\bibitem{carfora-ellis-94-p-a}
M~Carfora and G.~Ellis.
\newblock The geometry of classical change of signature.
\newblock {\em Preprint}, 1994.

\bibitem{hayward-93a}
S.~A. Hayward.
\newblock On cosmological isotropy, quantum cosmology and the weyl curvature
  hypothesis.
\newblock {\em Class. Quantum Grav.}, 10:L7--L11, 1993.

\bibitem{hayward-94-p-a}
S.~A. Hayward.
\newblock General laws of black-hole dynamics.
\newblock {\em Phys. Review D}, 49:6467--6474, 1994.

\bibitem{kossowski-kriele-93-p-a}
M.~Kossowski and M.~Kriele.
\newblock Transverse, type changing, pseudo {R}iemannian metrics with smooth
  curvature.
\newblock {\em Preprint FI93-CT11, The Fields Institute for Research in
  Mathematical Sciences}, 1993.

\bibitem{kossowski-85a}
M.~Kossowski.
\newblock Fold singularities in pseudo {R}iemannian geodesic tubes.
\newblock {\em Proc. Amer. Math. Soc.}, 95(3):463--469, 1985.

\bibitem{garabedian-86}
P.~R. Garabedian.
\newblock {\em Partial differential equations}.
\newblock Chelsea Publishing Company, New York, 2nd edition, 1986.



\end{thebibliography}

\end{document}